\documentclass{article}

\usepackage[a4paper, total={6.5in,9in}]{geometry}
\usepackage{graphicx, amsmath, tikz-cd, amsfonts, parskip,amsthm}
\newcommand{\pa} [2] {\frac{\partial #1}{\partial #2}} 

\newtheorem{theorem}{Theorem}[section]
\newtheorem{definition}{Definition}[section]

\begin{document}

\title {Extending the Kostant-Souriau quantization map by ``tautological-tuning" with symplectic and Riemannian structures}
\author{T. McClain \footnote{Department of Physics and Engineering, Washington and Lee University, Lexington, VA 24450 USA, email: mcclaint@wlu.edu}}

\maketitle

\small
Though quantization is -- from an ontological standpoint -- a very strange operation, it seems unavoidable in the actual practice of physics. From a mathematical standpoint, canonical quantization was superseded decades ago by more elegant constructions, yet among practicing physicists it remains the de-facto champion among the many alternatives. Despite this fact, there is to this day no mathematically well-defined, coordinate independent construction that reproduces the results of canonical quantization for the most physically important phase space functions: position, momentum, angular momentum, and (quadratic) Hamiltonian functions. In this paper, I construct such a quantization map by using standard structures from symplectic and Riemannian geometry in non-standard ways.

\normalsize
\textbf{Keywords:}
Differential Geometric Methods in Theoretical Physics, Frontiers in Mathematical Physics, Geometrical Methods in Mathematical Physics, Non-relativistic Quantum Mechanics 

\section{Introduction}

From a physical or ontological standpoint, the operation of quantization is a very strange one. After all, we believe that the classical, macroscopic world is made up out of more fundamental quantum, microscopic constituents, not the other way around. 

However, as far as the actual practice of physics is concerned, the operation of quantization seems to be unavoidable. Today, more than a hundred years after the birth of quantum theory, it is not only standard but ubiquitous to begin any analysis of a quantum theory -- whether of particles or of fields -- by considering a corresponding classical theory and then quantizing it using one of a large set of quantization schemes. In some ways, the sheer variety of these schemes is the best evidence of their essential nature. However, it is even clearer evidence that no single quantization scheme is universally applicable.

The earliest and most frequently used of these quantization schemes is surely canonical quantization. Though physically quite successful, canonical quantization is not a coordinate-independent mathematical process: given two different phase space coordinate systems $ X = \{ q^i, p_i \}$ and $Y = \{ Q^i, P_i \}$ and the standard position representation canonical quantization map $Q_C \mid  \{ q^i, p_i \} \mapsto \{ q^i, - \text{i} \hbar \frac{\partial }{\partial q_i} \} $, the following diagram does not commute

$$
\begin{tikzcd}
\{ q^i, p_i \} \arrow[r,"Q_C"] \arrow[d,"Y"] & \{q^i, - \text{i} \hbar \frac{\partial }{\partial q^i} \} \arrow[d,"Y"] \\
\{ Q^i, P_i \} \arrow[r,"Q_C"] & \{ Q^i, - \text{i} \hbar \frac{\partial }{\partial Q^i} \} \\
\end{tikzcd}
$$

Intuitively, this means that the result of canonical quantization depends explicitly on the coordinate system in which the quantization process is carried out.

This is by no means a new observation, and there have been decades of research into how best to solve the problem. The first attempts were mostly efforts to patch up the canonical quantization procedure by the addition of more ad hoc rules; see \cite{podolsky1928quantum} for perhaps the most famous and enduring of these. This line of research has continued all the way to the present day (see, for example, \cite{kleinert1997} and \cite{greiter2018landau}), and in fact remains more-or-less standard among experimentally minded quantum physicists. New discoveries of fundamental significance are still being made along these lines; see, for instance, \cite{muller2019neumann}.

However, among more mathematically minded quantum physicists (and, of course, among mathematicians) this ad hoc approach was long ago superseded by very different approaches devoted to solving the basic problem that the procedure of canonical quantization -- though physically very successful when implemented by sophisticated practitioners -- is nevertheless mathematically ill defined. There are two main approaches, the first being to create a proper functional calculus to formalize the ideas of quantization implicit in the canonical procedure; this path leads to things like Weyl quantization. The other path is to try to geometrize the quantization process, and there are two main research programs that take this approach. The first and foremost is geometric quantization, really begun by van Hove in the 1940s, but taken in its modern direction by Souriau and Kostant (among others) in the 1960s and 70s; see \cite{souriau1966quantification} and \cite{kostant1970quantization}. The second is deformation quantization, also really begun by van Hove, but given its modern form by Kontsevich (among others); see \cite{kontsevich2003deformation}. Both research programs can claim some major successes: mostly mathematical in the case of deformation quantization, both mathematical and physical in the case of geometric quantization. Researchers continue to produce new results of fundamental interest in geometric \cite{tuynman2016metaplectic} \cite{prieto2017geometry} and deformation \cite{lechner2016strict} \cite{calaque2017shifted} quantization, as well as to find new applications of the mathematical techniques of each program (see, for example, \cite{galasso2016remarks} and \cite{karabegov2019formal}).

However, all of these quantization programs must contend with a fundamental problem: the theorems of Groenewald \cite{groenewold1946principles} and Van Hove \cite{van1951problem} (among others; see, for example, \cite{gotay1996obstruction}) make it clear that no quantization map can satisfy all the requirements one might hope to impose upon it for all possible functions on phase space. 

The goal of this paper is to introduce a geometric approach to quantization that reproduces the results of canonical quantization\footnote{Or, in the case of quadratic Hamiltonians, the correct extension of canonical quantization to curved spaces.} in a coordinate-independent manner using only standard symplectic and Riemannian structures of the base and phase space manifolds. In light of the Groenewald-Van Hove theorems, no effort is made to make sure that the quantization map ``works" for all possible functions on phase space. Instead, this approach assumes that the map only needs to quantize a tiny handful of important phase space functions to be judged physically successful. I take it that the ``natural" phase space functions to consider in the non-relativistic case are the position coordinate functions and the infinitesimal generators of spacetime translations, namely the Hamiltonian and momentum functions. In the especially important case of three-dimensions, one can reasonably argue that the angular momentum functions should be included as well. (See, for instance, the classical text \cite{abraham1978foundations} for a more thorough account of how and why these are appropriate phase space functions to consider.) 

Admittedly, it would be better to have a rigorous way of defining which functions are necessary to quantize rather than to say that they should be ``physically important." But my hope is that this work will provide an interesting alternative approach to the quantization of particle systems that will appeal to researchers regardless of their personal preferences with regard to which phase space functions should be regarded as ``physically important."

To summarize, the main assumptions that motivate this particular quantization scheme are that

\begin{enumerate}
\item The need for quantization schemes in general seems unavoidable
\item We would like to find a quantization map that mimics the major successes of canonical quantization 
\item We would like that map to be mathematically well-defined and coordinate independent
\item We only need the map to successfully quantize a small handful of phase space functions, and indeed can only critically evaluate such a map on those few functions
\end{enumerate}

More will be said about the ultimate merits and demerits of this approach in the conclusion. 

The outline of the paper is as follows. I review the basics of symplectic geometry in Section \ref{symplecticstructures}, in particular those structures that will be useful in constructing the quantization map. Section \ref{ksquantization} reviews the simplest construction of the Kostant-Souriau quantization map that was the jumping-off point for the program of geometric quantization and that I take in a rather different direction here, while in Section \ref{projected} I modify this map in a non-standard way that allows it to produce physically more successful results at the cost of a certain degree of mathematical elegance. In Sections \ref{connections}, \ref{tautological}, and \ref{hessian} I build up the rest of the essential geometric tools, most importantly some useful definitions of generalized, coordinate independent notions of constant-, linear-, and quadratic-in-momentum phase space functions in Definitions \ref{def:f0}, \ref{def:f1}, and \ref{def:f2} that allow the construction of the tautological tuning functions in Theorems \ref{thm:x1} and \ref{thm:x2}, whose kernels are (linear combinations of) these generalized constant-, linear-, and quadratic-in-momentum phase space functions. I then use these tools in Section \ref{ttquantization} to define the quantization map \eqref{ttq}, about which I then prove the following theorem:

\textbf{Theorem 8.1} \emph{Given an arbitrary metric $g$ on $P$ and a connection $\nabla$ on $TP$ that satisfies the condition $d \pi( \nabla_v u) = 0$ for all vertical vectors $v \in VP$ and all vector fields $u$ that satisfy $d \pi ([v, u]) = 0$ for all $v \in VP$, the tautologically-tuned quantization map of \eqref{ttq} has the following three properties:
\begin{enumerate} 
\item It is well-defined, independent of coordinates
\item It reduces to the Kostant-Souriau quantization map for functions $f$ such that $f \in F_0(P)$ or $f \in F_1(P)$, as defined in Definitions \ref{def:f0} and \ref{def:f1}
\item It correctly quantizes quadratic Hamiltonian functions $H = \frac{1}{2m} g^{i j}(q^i) p_i p_j + V(q^i)$ in arbitrary coordinate systems
\end{enumerate}}

Finally, I discuss the merits and demerits of the tautologically-tuned quantization scheme and directions for future research in Section \ref{conclusions}. 

\section{Symplectic structures for quantization} \label{symplecticstructures}

The material in this section is standard and can be found in any textbook on geometric Hamiltonian mechanics or symplectic geometry (see, for example, \cite{abraham1978foundations}). Readers already familiar with symplectic geometry can safely skip this section, as every effort has been made to make sure that it aligns with the notational conventions most common in the field. For readers at the opposite end of the spectrum, a certain amount of differential topology is necessary to understand symplectic geometry. The necessary material concerning differential manifolds, tangent spaces, differential forms, etc. can be found in any textbook on differential topology, as well as a good many textbooks on geometric methods in physics (see, for example, \cite{schutz1980geometrical}).

Let the differentiable manifold $Q$ represent the space in which our particle or particles are able to move. In the most common case, this is simply three-dimensional space $Q = \mathbb{R}^3$. The phase space for the particle is then
\begin{equation}
P = T^*Q
\end{equation}
with projection map $\pi : P \to Q$. It is this space upon which all the geometric structure of the theory will be built. The fundamental element of the symplectic structure of this space is the tautological (or canonical, or many other names) one-form, $\theta$. It lives not on $P$ but on the cotangent space $T^*P$, and it is defined intrinsically by
\begin{equation}
\theta_p(v) = p \circ d \pi (v)
\end{equation}
where $v \in T_p P$ is any vector in the tangent space $TP$ over the point $p$, $d \pi : TP \to TQ$ is the differential of the projection map $\pi$, and $p \in P$ is any point in $P$. 

Local fibered coordinates on $P$ are coordinates $\{ q^i, p_i \}$ on $P$ such that there exist coordinates $ \{ x^i \} $ on $Q$ with $q^i(p) = x^i(\pi(p))$ for all $p$ in the domain of the coordinate chart. These local fibered coordinates naturally induce coordinates on $TP$ in which the basis vectors for the fibers $T_p P$ are given by the coordinate derivatives $\pa{}{q^i}$ and $\pa{}{p_i}$ evaluated at each point $p$ in the domain of the chart. These tangent bundle coordinates are also called local fibered coordinates, and induce compatible coordinates on the dual bundle $T^*P$ in the usual way. 

In local fibered coordinates one can write $v \in T_p P = v^i \frac{\partial }{\partial q^i} + v_i \frac{\partial }{\partial p_i}$, $ d \pi = \frac{\partial }{\partial q^i} \otimes dq^i$, and $p = p_i dq^i + q^i e_i$ so that $\theta_p(v) = v^i p_i $. In other words, $\theta$ is a one-form on $T^*P$ that can be written in local fibered coordinates as
\begin{equation}
\theta = p_i dq^i
\label{intrinsictheta}
\end{equation}

Though the typical use of the tautological one-form $\theta$ is simply to produce the symplectic form $\omega$, my approach to quantization will make more use of it than is standard. Indeed, it is the word tautological from the name tautological one-form that gives rise to the name tautologically-tuned quantization, for reasons that should soon become clear.

The single most important symplectic structure in the standard approach is undoubtedly the symplectic form $\omega$, the exterior derivative of the tautological one-form
\begin{equation} \label{symplecticform}
\omega = d \theta = dp_i \wedge dq^i 
\end{equation}
where the second equality holds in (and indeed defines) canonical coordinates on the manifold $P$. In the case that $M = \mathbb{R}^3$, this reads (for those unfamiliar with the Einstein summation convention and/or wedge product)
\begin{equation}
\omega = dp_1 \otimes dq^1 +  dp_2 \otimes dq^2 +  dp_3 \otimes dq^3 - dq^1 \otimes dp_1 - dq^2 \otimes dp_2 - dq^3 \otimes dp_3
\end{equation}
Since the symplectic form is non-degenerate (meaning that $\omega(u,v) = 0 \ \forall \ v \iff u = 0$), it is possible to associate with each function $f \in C^\infty(M)$ a vector field $X_f \in X(M)$ (usually called the Hamiltonian vector field of $f$) via the requirement
\begin{equation}
\omega(X_f, -) = df
\end{equation}
In local canonical coordinates, this amounts to the assignment
\begin{equation}
X_f = - \frac{\partial f}{\partial p_i} \frac{\partial }{\partial q^i} + \frac{\partial f}{\partial q^i} \frac{\partial }{\partial p^i}
\end{equation}
This assignment in turn makes it possible to define a Poisson structure $\Pi$, with the defining property that
\begin{equation}
\Pi (df, -) = X_f
\end{equation}
for all $f \in C^\infty(M)$. This definition of $\Pi$ gives us the canonical coordinate description
\begin{equation} \label{Pi}
\Pi = \frac{\partial }{\partial q^i} \wedge \frac{\partial }{ \partial p_i}
\end{equation}
Finally, it is necessary to define a less standard symplectic structure, namely the vector field that results from contracting the Poisson structure $\Pi$ and the tautological one-form $\theta$. One might call this the tautological vector field. Though it is not the Hamiltonian vector field of any function $f$, it is essential in tautologically-tuned quantization. In analogy with the Hamiltonian vector fields, I will call it $X_\theta$. The operations defined above give us
\begin{equation} \label{chitheta}
X_\theta = \Pi(\theta) = p_i \frac{\partial }{\partial p_i}
\end{equation}
where the second equality once again holds in local canonical coordinates on $P$.
 
\section{The Kostant-Souriau quantization map} \label{ksquantization}
There exists a simple quantization scheme that uses only the symplectic structures of the previous section to produce a map from smooth functions on the phase space $P$ to linear operators on (complex) phase space functions. This map -- which was the starting point for the geometric approach to quantization introduced by Kostant and Souriau in the 1970s -- is given by
\begin{equation} \label{ksoriginal}
Q_{KS}(f) := f - X_\theta f + \text{i} \hbar X_f = f - p_i \frac{\partial f}{\partial p_i} + \text{i} \hbar \frac{\partial f}{\partial q^i} \frac{\partial }{\partial p_i} - \text{i} \hbar \frac{\partial f}{\partial p_i} \frac{\partial}{\partial q^i}
\end{equation}
Though relatively straightforward to define, this quantization map has several very nice properties. For instance, it maps the canonical coordinate functions to almost appropriate looking operators:
\begin{equation}
Q_{KS}(q^i) = q^i + \text{i} \hbar \frac{\partial }{\partial p_i}
\end{equation}
(but note the strange looking momentum coordinate derivative) and
\begin{equation}
Q_{KS}(p_i) = - \text{i} \hbar \frac{\partial }{\partial q^i}
\end{equation}
It even maps the angular momentum functions to almost appropriate looking operators:
\begin{equation}
Q_{KS}(L_1) = Q_{KS}(q^2 p_3 - q^3 p_2) = \text{i} \hbar \left( p_3 \frac{\partial }{\partial p_2} - p_2 \frac{\partial }{\partial p_3} - q^2 \frac{\partial }{\partial q^3} + q^3 \frac{\partial }{\partial q^2} \right)
\end{equation}
\begin{equation}
Q_{KS}(L_2) = Q_{KS}(q^3 p_1 - q^1 p_3) = \text{i} \hbar \left( p_1 \frac{\partial }{\partial p_3} - p_3 \frac{\partial }{\partial p_1} - q^3 \frac{\partial }{\partial q^1} + q^1 \frac{\partial }{\partial q^3} \right)
\end{equation}
\begin{equation}
Q_{KS}(L_3) = Q_{KS}(q^1 p_2 - q^2 p_1) = \text{i} \hbar \left( p_2 \frac{\partial }{\partial p_1} - p_1 \frac{\partial }{\partial p_2} - q^1 \frac{\partial }{\partial q^2} + q^2 \frac{\partial }{\partial q^1} \right)
\end{equation}
The presence of the momentum coordinate derivatives in these operators is embarrassing, but if these could be eliminated then the operators would match the results of canonical quantization. However, even if we assume that we can fix this particular issue relatively easily (which we can, as we will see in the next section), it is easy to see that this map is still far from perfect by looking at how it fails to correctly quantize typical Hamiltonian operators. For example, the one dimensional simple harmonic oscillator Hamiltonian maps to 
\begin{equation} \label{kssho}
Q_{KS}(H_{SHO}) = Q_{KS} \left( \frac{p^2}{2m} + \frac{1}{2} m \omega^2 q^2 \right) = \text{i} \hbar \left(- \frac{p}{m} \frac{\partial }{\partial q} + m \omega^2 q \frac{\partial }{\partial p} \right) - \frac{p^2}{2m} + \frac{1}{2} m \omega^2 q^2
\end{equation}
which is a far cry from the expected
\begin{equation}
Q(H_{SHO}) = - \frac{\hbar^2}{2m} \frac{\partial ^2}{\partial q^2} + \frac{1}{2} m \omega^2 q^2
\end{equation}
of canonical quantization and correct physics.

Solving this problem with Hamiltonian operators will be the main achievement of the tautologically-tuned quantization map. 

\section{The projected Kostant-Souriau map and the space of quantum states} \label{projected}

From a physical perspective, there are two major problems with the Kostant-Souriau quantization map considerably more fundamental than the fact that it does not correctly quantize Hamiltonian functions quadratic in the momentum coordinates. First, as already noted, there are extraneous derivatives in these operators; if we choose the position representation for our quantum states, these are the momentum derivatives. Second, these operators naturally act on the space of functions on the full phase space $P$, while the actual space of quantum states in the position representation should be only functions over the base manifold $Q$. Though it is not common in geometric quantization, we can solve both of these problems using a natural structure of the phase space manifold $P = T^*Q$, namely the projection map $\pi : P \to Q$.

The simple solution to the first of these problems is to project the vector parts of our operators from $TP$ to $TQ$ using the differential of the projection map $d \pi : TP \to TQ$. In local fibered coordinates in which $d \pi = \frac{\partial}{\partial q^i} \otimes dq^i + 0 \frac{\partial}{\partial p_i} \otimes dp_i $, doing so gives us:

\begin{equation} \label{ksprojected} Q_{PKS}(f) =  f - X_\theta f + \text{i} \hbar \, d\pi( X_f ) = f - p_i \frac{\partial f}{\partial p_i} - \text{i} \hbar \frac{\partial f}{\partial p_i} \frac{\partial}{\partial q^i} \end{equation}

which immediately eliminates all the momentum derivatives in the operators:

\begin{equation}
Q_{PKS}(q^i) = q^i
\end{equation}

\begin{equation}
Q_{PKS}(p_i) = - \text{i} \hbar \frac{\partial }{\partial q^i}
\end{equation}

\begin{equation}
Q_{PKS}(L_1) = Q_{PKS}(q^2 p_3 - q^3 p_2) = \text{i} \hbar \left(q^3 \frac{\partial }{\partial q^2} - q^2 \frac{\partial }{\partial q^3} \right)
\end{equation}

\begin{equation}
Q_{PKS}(L_2) = Q_{PKS}(q^3 p_1 - q^1 p_3) = \text{i} \hbar \left(q^1 \frac{\partial }{\partial q^3} - q^3 \frac{\partial }{\partial q^1} \right)
\end{equation}

\begin{equation}
Q_{PKS}(L_3) = Q_{PKS}(q^1 p_2 - q^2 p_1) = \text{i} \hbar \left(q^2 \frac{\partial }{\partial q^1} - q^1 \frac{\partial }{\partial q^2} \right)
\end{equation}

which all now match the results of canonical quantization exactly.

Unfortunately, the projected Kostant-Souriau map does not solve the problem with quadratic Hamiltonians, as we have

\begin{equation}
Q_{PKS}(H_{SHO}) = Q_{PKS} \left( \frac{p^2}{2m} + \frac{1}{2} m \omega^2 q^2 \right) = - \text{i} \hbar \frac{p}{m} \frac{\partial }{\partial q} - \frac{p^2}{2m} + \frac{1}{2} m \omega^2 q^2
\end{equation}

which is almost as bad as before.

Taking this easy route also allows us -- indeed, requires us -- to stipulate that proper physical wavefunctions are functions only on $Q$, not on $P$.

There are two substantial mathematical drawbacks to this approach that are responsible for its relative lack of popularity. The first is that, since the projection operation requires us to consider only physically appropriate wavefunctions in the position representation, we can no longer use the natural symplectic volume form $\text{vol}_P =  \omega^{\dim Q}$ (where the product implicit in the exponentiation is the exterior product), as it only exists on the full phase space $P$. Therefore, to define an inner product -- which in turn will allow us to form a proper Hilbert space, calculate expectation values, etc. -- we need more structure, specifically a Riemannian metric on $Q$. 

Furthermore, while $X_f$ is a true vector field over $P$, $d \pi(X_f)$ is not generically a vector field on $TQ$: while each vector is defined point-by-point, the various vectors need not combine to create a proper vector field on $TQ$ because the map $\pi: P \to Q$ is not injective. Only in the case that we are able to carefully eliminate all the dependence of the resulting derivative operators on the momenta will these projections amount to true vector fields on $TQ$. Note, however, that if we are able to fully implement something like canonical quantization, then this process does yield derivative operators that are well-defined vector fields on $TQ$, as in the case of the functions $q^i$, $p_i$, $L_i$ considered above.

So, from the perspective of canonical quantization, the biggest remaining problem with the projected Kostant-Souriau quantization map is that it fails to correctly quantize quadratic Hamiltonian functions. Fixing this problem will require a few extended asides on mathematics.

\section{Hamiltonian functions, metrics, and connections} \label{connections}

Defining an inner product on our space of functions on $Q$ to form a proper Hilbert space can be accomplished by introducing a metric on $Q$. We will see in Section \ref{ttquantization} that defining a tautologically-tuned quantization map also requires that we introduce a covariant derivative (connection) on the space of vector fields over $P$. In this section, I will discuss how and why these two requirements are in fact natural from the perspective of the phase space functions we still need to quantize.

The most physically important functions that we cannot yet correctly quantize are the single particle Hamiltonian functions, with the general form

\begin{equation} \label{hamiltonian} H = T + V = \frac{1}{2m} g^{i j} p_i p_j + V(q^i) \end{equation}

Defining the kinetic part of these functions in a coordinate-invariant way almost invariably relies on introducing a metric on the fibers of $P = T^*Q$.\footnote{This is true even if these metrics are simply canonical flat metrics; the natural way to write the kinetic term in these cases is $T = \delta^{ i j} p_i p_j$.} It is worth pointing out that in standard cases like those of the harmonic oscillator, gravitational, or electric potential energy functions, we also require a metric on $Q$ itself. However, since this is not universal (for example, if we are only considering a free particle), let us not directly assume this second metric structure. In any case, since a metric on $T^*Q$ immediately induces a metric on $TQ$ (and vice versa), our metric on $T^*Q$ immediately eliminates one of the main drawbacks of using the projection operator $d \pi : TP \to TQ$ in our quantization maps: namely that we did not originally have any natural Riemannian metric on $Q$.

As a brief reminder, a metric $g$ on $TQ$ defines a unique covariant derivative $\nabla_{TQ}$ called the Levi-Civita connection such that $ \nabla g = 0 $ and $\nabla X(Y) - \nabla Y(X) = 0$.\footnote{In this section and throughout the rest of the work, I will take the metric on $TQ$ to be the fundamental one because this accords more closely with the standard in Riemannian geometry, even though the metric on $P = T^*Q$ is actually the one we get from the physics of the situation.} In coordinates on $TQ$ in which $g = g_{i j} dx^i \otimes dx^j$ and given any vector $v = v^i \frac{\partial}{\partial x^i} \in TQ$ and one-form $\alpha = \alpha_i dx^i \in P = T^*Q$, these requirements give us

$$ \nabla_{TQ} (v) = \left( \frac{\partial v^i}{\partial x^j} + v^k \Gamma^i_{k j} \right) \, dx^j \otimes \frac{\partial}{\partial x^i} $$

and

$$ \nabla_P (\alpha) = \left( \frac{\partial \alpha_i}{\partial x^j} - \alpha_k \Gamma^k_{j i} \right) \, dx^j \otimes d x^i $$

where the Christoffel symbols $\Gamma^i_{j k}$ are defined by

$$ \Gamma^i_{j k} = \frac{1}{2} g^{i l} \left( \frac{\partial g_{l j}}{\partial x^k} +  \frac{\partial g_{k l}}{\partial x^j} -  \frac{\partial g_{j k}}{\partial x^l} \right) $$

The foregoing is all standard Riemannian geometry. However, the connection we will end up needing to define a successful tautologically-tuned quantization map is a connection on $TP = T(T^*Q)$, not on $P = T^*Q$ or $TQ$ as immediately given by our Riemannian metric. Unfortunately, there is no simple way to build this connection using the structures already given. Some ideas about how one might build this this connection in a somewhat roundabout way will be given in Section \ref{conclusions}. 

\section{Structures for tautological tuning} \label{tautological}

Though the tautological tuning process will use only the standard structures from symplectic geometry referenced in Section \ref{symplecticstructures}, it will use them in non-standard combinations that are worth precisely identifying and naming.

Before diving into the details, it might be helpful to give the intuitive reasoning behind the mathematics. The intuition is that to mimic the process of canonical quantization using natural geometric structures, it seems necessary to treat different phase space functions differently. This is only possible within a unified geometric framework if we can come up with a coordinate-invariant way to identify the different kinds of phases space functions that we wish to treat differently. More specifically, we would like to come up with a way to identify -- in an abstract, coordinate-invariant way -- phase space functions that are 1) constant 2) linear and 3) quadratic in the momentum variables naturally picked out by the projection map $\pi: P \to Q$. In this section, I will detail some of many ways to do this by defining coordinate-invariant maps whose kernels are (linear combinations of) these sets of phase space functions.

To do this, we begin with the vector field $X_\theta$ defined in Section \ref{symplecticstructures} that will form the backbone of the tautological tuning process:

\begin{equation}
X_\theta = \Pi(\theta) = p_i \frac{\partial }{\partial p_i}
\end{equation}

where as usual the second equality holds in local canonical coordinates. This leads us to the following definition:

\begin{definition} \label{def:f0}

The function

\begin{equation} \Delta_0 f := X_\theta(f) =p_i \frac{\partial f}{\partial p_i}  \label{delta0} \end{equation}

naturally serves to define the set 

\begin{equation} F_0(P) := \ker \Delta_0 \end{equation}

of functions that do not depend on the phase space variables $p_i$ naturally picked out by the projection map $\pi: P \to Q$.

\end{definition}

This simple, coordinate-invariant notion of constancy suggests the following natural extension:

\begin{definition} \label{def:f1}

The function

\begin{equation} \Delta_1 f := f - X_\theta(f) = f - p_i \frac{\partial f}{\partial p_i}  \label{delta1} \end{equation}

naturally\footnote{By ``naturally," I mean that this is the only function $f + \alpha \chi_\theta $ whose kernel is the set of linear functions that we want.} serves to define the set 

\begin{equation} F_1(P) := \ker \Delta_1 \end{equation}

of functions linear in the phase space variables $p_i$ naturally picked out by the projection map $\pi: P \to Q$.

\end{definition}

If $\dim Q = 1$ the equation $ \Delta_1 = 0 $ has the unique solution $f = g(q) p$. Naturally, this equation has more solutions when $ \dim Q > 1$; for example, when $\dim Q = 3$, another solution is $f = p_1 + 2 p_2 + 3 p_3$. However, it is still the case that this equation serves to pick out functions that are linear in the momentum variables from all the others.

These simple insights serve as the gateway to tautological tuning and the following theorem:

\begin{theorem} \label{thm:x1}
The function

\begin{equation} \chi_1 f := \lim_{\epsilon \rightarrow 0} \frac{\epsilon}{\epsilon + \Delta_1 f} \label{chi1} \end{equation}

is one when $f \in F_1(P)$ and zero when it is not. 
\end{theorem}

\begin{proof}
Since $F_1(P) = \ker \Delta_1$ by definition, this simply amounts to saying that the given limit is $1$ when $\Delta_1 f  = 0$ and $0$ when $\Delta_1 f \neq 0$.
\end{proof}
 
This map will ultimately allow us to pick out linear phase space functions for appropriate treatment in the quantization process.

Since we are interested not only in quantizing phase space functions that are linear in the momentum variables -- which is already done quite readily with the Kostant-Souriau quantization map -- but also those that are quadratic in the momentum variables, the natural generalization of the previous definition would seem to be to define the function 

$$ f - \frac{1}{8} X^2_\theta(f) = f - \frac{1}{8} p_i \frac{\partial f}{\partial p_i} - \frac{1}{8} p_i p_j \frac{\partial^2 f}{\partial p_i \partial p_j} $$

whose kernel would serve to define\footnote{In this case, I have omitted the word ``naturally" from this construction, as there are infinitely many combinations $f + \alpha \chi_\theta + \beta \chi^2_\theta$ with the same kernel. I have arbitrarily chose the one with $\alpha =0$.} the set of functions quadratic in the phase space variables $p_i$ naturally picked out by the projection map $\pi: P \to Q$. If our base manifold $Q$ were one-dimensional, this defining equation has the unique solution $f = g(q) p^2$. Again, this equation has more solutions when $\dim Q > 1$, such as $f = 5 p_1 p_2$, but it would still serve to pick out functions that are quadratic in (combinations of) the momentum variables.

Unfortunately, the physics of the situation is not quite so elegant as the mathematics. While the only physically important phase space functions linear in the momenta (to whit, $p_i$ and $L_i$) contain no constant-in-momentum terms, the physically important phase space functions quadratic in the momenta are the Hamiltonian functions $H = \frac{1}{2m} g^{i j} p_i p_j + V(q^i)$, which naturally contain the constant-in-momentum terms $V(q^i)$. Therefore, the more important tautological tuning function comes from the following:

\begin{definition} \label{def:f2}

The function

\begin{equation} \Delta_2 f := X_\theta(f) - \frac{1}{2} X^2_\theta(f) = \frac{1}{2} p_i \frac{\partial f}{\partial p_i} - \frac{1}{2} p_i p_j \frac{\partial^2 f}{\partial p_i \partial p_j} \label{delta2} \end{equation}

serves to define the set

\begin{equation} F_2(P) := \ker \Delta_2 \end{equation}

of functions that are linear combinations of 1) constant and 2) quadratic in the phase space variables $p_i$ naturally picked out by the projection map $\pi: P \to Q$.

\end{definition}

If our base manifold $Q$ is one-dimensional, the equation $ \Delta_2 = 0 $ has the unique solution $f = h(q) + g(q) p^2$. Again, this equation has more solutions when $\dim Q > 1$, such as $f = 5 p_1 p_2 - q^3$, but it still serves to pick out functions that are linear combinations of functions that constant or quadratic in (combinations of) the momentum variables. In particular, we have the following:

\begin{theorem} \label{thm:x2}

The function

\begin{equation} \chi_2 f := \lim_{\epsilon \rightarrow 0} \frac{\epsilon}{\epsilon + \Delta_2 f} \label{chi2} \end{equation}

is one when $f \in F_2(P)$, and zero when it is not.
\end{theorem}

\begin{proof}
Since $F_2(P) = \ker \Delta_2$ by definition, this amounts to saying that the given limit is $1$ when $\Delta_2 f  = 0$ and $0$ when $\Delta_2 f \neq 0$.
\end{proof}

This analysis of the quadratic case shows that these tautological tuning functions are by no means unique, or even necessarily natural in any deep way. Indeed, the question of whether there are other tautological tuning functions which might eliminate the need for projecting the Kostant-Souriau quantization map -- or provide other interesting benefits -- is a natural subject for future work.

\section{From tensor products to multiple derivative operators} \label{hessian}

The ultimate output of any quantization map built from symplectic or Riemmanian structures will be some kind of tensor. However, the canonical quantization process naturally produces higher derivative operators when applied to functions that are quadratic in the momentum variables. The problem, as is well known, is that multiple derivative operators are not naturally tensorial. For example, a second derivative operator does not transform as a rank two tensor, meaning that the following diagram does not commute

$$
\begin{tikzcd}
T^{i j} \frac{\partial}{\partial x^i} \otimes \frac{\partial}{\partial x^j} \arrow[r,"X \to \tilde X"] \arrow[d,"\text{SD}"] & \tilde T^{i j} \frac{\partial}{\partial \tilde x^i} \otimes \frac{\partial }{\partial \tilde x^i} \arrow[d,"\text{SD}"] \\
T^{i j} \frac{\partial^2}{\partial x^i \partial x^j} \arrow[r,"X \to \tilde X"] & \tilde T^{i j} \frac{\partial^2}{\partial \tilde x^i \partial \tilde x^j} \\
\end{tikzcd}
$$

where the mapping SD is the simplest possible way to form a second derivative from the tensor product of two vector fields: $\text{SD}(T^{i j} \frac{\partial}{\partial x^i} \otimes \frac{\partial}{\partial x^j} ) = T^{i j} \frac{\partial^2}{\partial x^i \partial x^j}$.

This non-commutativity occurs because of the action of one of the partial derivatives in the composition on the coordinate transformation of the other partial derivative. That is

$$ u^i v^j \frac{\partial^2}{\partial x^i \partial x^j} = \frac{\partial x^i}{\partial \tilde x^l} \tilde u^l \frac{\partial x^j}{\partial \tilde x^m} \tilde v^m \frac{\partial \tilde x^n}{\partial x^i} \frac{\partial}{\partial \tilde x^n} \left( \frac{\partial \tilde x^o}{\partial x^j} \frac{\partial}{\partial \tilde x^o} \right) = \tilde u^l \tilde v^m \frac{\partial^2}{\partial \tilde x^l \partial \tilde x^m} + \tilde u^l \tilde v^m \frac{\partial x^j}{\partial \tilde x^m} \frac{\partial^2 \tilde x^o}{\partial \tilde x^n \partial x^j} \frac{\partial}{\partial \tilde x^o} $$

which only allows our diagram to commute when the term $ \frac{\partial^2 \tilde x^o}{\partial \tilde x^n \partial x^j} = 0$.\footnote{There are restricted categories of coordinate transformations for which this term is indeed zero. One example is linear transformations, for which $\frac{\partial \tilde x^o}{\partial x^j}$ is constant so that $\frac{\partial}{\partial \tilde x^n} \left( \frac{\partial \tilde x^o}{\partial x^j} \right) = 0$.}

There is a well-known solution to this problem: while bare second derivatives are not tensorial, the Hessian tensor is:

\begin{equation} \label{hessiantensor} \text{Hess}(f) = \nabla^2 f = \nabla df \end{equation}

where the definition requires the introduction of a connection $\nabla$ and the second equality holds because every connection reduces to the exterior derivative when acting on functions. Since they will be important for us, the components of the Hessian tensor are given by 

\begin{equation} \label{hessiancomp} \text{Hess}_{i j} = \nabla^2_{i j} := \nabla_i \nabla_j - \nabla_{\nabla_i (\frac{\partial}{\partial x^j})} \end{equation}

When acting on functions, the components of the Hessian tensor are simply

\begin{equation} \label{hessiancomp2} \text{Hess}_{i j} = \frac{\partial^2}{\partial x^i \partial x^j} - \Gamma^k_{i j} \frac{\partial}{\partial x^k} \end{equation}

where the $\Gamma^k_{i j}$ are the Christoffel symbols associated with the connection $\nabla$ in the chosen coordinate system.

The critical utility of the Hessian tensor in tautologically-tuned quantization is that the following tensor-to-second-derivative diagram does commute

$$
\begin{tikzcd}
T^{i j} \frac{\partial}{\partial x^i} \otimes \frac{\partial}{\partial x^j} \arrow[r,"X \to \tilde X"] \arrow[d,"\text{CSD}"] & \tilde T^{i j} \frac{\partial}{\partial \tilde x^i} \otimes \frac{\partial }{\partial \tilde x^i} \arrow[d,"\text{CSD}"] \\
T^{i j} \frac{\partial^2}{\partial x^i \partial x^j} -  T^{i j} \Gamma^k_{i j} \frac{\partial}{\partial x^k} \arrow[r,"X \to \tilde X"] & \tilde T^{i j} \frac{\partial^2}{\partial \tilde x^i \partial \tilde x^j} - \tilde T^{i j} \tilde \Gamma^k_{i j} \frac{\partial}{\partial \tilde x^k} \\
\end{tikzcd}
$$

when the mapping CSD is defined using the components of the Hessian tensor via 

\begin{equation} \label{csd} \text{CSD}(T^{i j} \frac{\partial}{\partial x^i} \otimes \frac{\partial}{\partial x^j} ) = T^{i j} \, \text{Hess}_{i j}= T^{i j} \frac{\partial^2}{\partial x^i \partial x^j} - T^{i j} \Gamma^k_{i j} \frac{\partial}{\partial x^k} \end{equation}

\section{Tautologically-tuned quantization} \label{ttquantization}

With these structures in place, it is time to define the tautologically-tuned quantization map that is the fundamental new construction in this paper:

\begin{equation} \label{ttq} Q_{TT}(f) := f + \chi_1 f \left( d \pi \circ \left( \text{i} \hbar \, \Pi \circ \nabla \right) f - X_\theta f \right) + \frac{1}{2} \chi_2 f \left( \text{CSD} \circ d \pi \circ \left( \text{i} \hbar \, \Pi \circ \, \nabla \right)^2 f - X_\theta f \right) \end{equation}

where $\chi_1$ and $\chi_2$ are the tautological tuning functions of \eqref{chi1} and \eqref{chi2}, $d \pi : TP \to TQ$ is the differential of the projection map $\pi : P \to Q$ used in Section \ref{projected}, $\Pi$ is the Poisson structure of \eqref{Pi}, $X_\theta$ is the tautological vector field of \eqref{chitheta}, CSD is the tensor-to-second-derivative map defined by \eqref{csd}, and $\nabla$ is an affine connection on $TP$ that constitutes a genuinely new piece of structure.\footnote{Note that since the first application of this covariant derivative is on an ordinary function, there is no need to choose any special connection for the $\chi_1$ term; all connections act the same way on functions. I write it this way only to emphasize the similarities between the two terms.}

This definition allows me to state the following theorem, which constitutes the main result of the paper:

\begin{theorem} \label{thm:ttq}
Given an arbitrary metric $g$ on $P$ and a connection $\nabla$ on $TP$ that satisfies the condition $d \pi( \nabla_v u) = 0$ for all vertical vectors $v \in VP$ and all vector fields $u$ that satisfy $d \pi ([v, u]) = 0$ for all $v \in VP$, the tautologically-tuned quantization map of \eqref{ttq} has the following three properties:
\begin{enumerate} 
\item It is well-defined, independent of coordinates
\item It reduces to the Kostant-Souriau quantization map for functions $f$ such that $f \in F_0(P)$ or $f \in F_1(P)$, as defined in Definitions \ref{def:f0} and \ref{def:f1}
\item It correctly quantizes quadratic Hamiltonian functions $H = \frac{1}{2m} g^{i j}(q^i) p_i p_j + V(q^i)$ in arbitrary coordinate systems
\end{enumerate}
\end{theorem}

\begin{proof}
The first criteria is met by construction, since the quantization map is formed by the successive application of coordinate-independent, differential geometric maps. 

If $f \in F_0(P)$, then $\chi_1(f) = 0$. While $\chi_2(f) = 1$ in this case, $d\pi \circ i \hbar \, \Pi \circ \nabla f = 0 = X_\theta f$, so $Q_{TT}(f) = f$, just as in the Kostant-Souriau quantization map. If $f \in F_1(P)$, then $\chi_2(f) = 0$ and $\chi_1(f) = 1$, so the map reduces to precisely the Kostant-Souriau prescription. So the second criteria is also met.

The proof the final criteria is more involved. In particular, it is here that the connection on $TP$ with the property stated above becomes necessary.

As a first step, we note that if $f = H = \frac{1}{2m} g^{i j}(q^i) p_i p_j + V(q^i)$, then $\chi_1(f) = 0$ and $\chi_2(f) =1$, so that $Q_{TT}$ reduces to $f + \frac{1}{2} \left( \text{CSD} \circ d \pi \circ \left( \text{i} \hbar \, \Pi \circ \, \nabla \right)^2 f - X_\theta f \right)$. Also, $\frac{1}{2} X_\theta f = \frac{1}{2m} g^{i j} p_i p_j $, so that $f - \frac{1}{2}X_\theta f = V(q^i)$. 

It remains only to tackle the term $\frac{1}{2} \text{CSD} \circ d \pi \circ \left( \text{i} \hbar \, \Pi \circ \, \nabla \right)^2 f$. We will at first take the affine connection $\nabla$ to be arbitrary; we will find that the non-trivial requirement on $\nabla$ to allow our map to correctly quantize quadratic Hamiltonians on arbitrarily curved spaces is precisely the one stated above. In this case, $X_f = \Pi(\nabla f, -) = - \frac{g^{i j}}{m} p_j \pa{}{q^i} + \frac{\partial V}{\partial q^i} \frac{\partial}{\partial p_i}$, so we can calculate

\begin{multline*} \nabla(X_f) = \left(\pa{V}{q^k} G^{i k}_j - \frac{1}{m} \pa{g^{ik}}{q^j} p_k - \frac{1}{m}  g^{k l} p_l A^i_{j k} \right) dq^j \otimes \pa{}{q^i} + \left(\pa{^2 V}{q^i \partial q^j} + \pa{V}{q^k} E^k_{i j} - \frac{1}{m} g^{kl} p_l C_{i j k} \right) dq^j \otimes \pa{}{p_i} \\
 + \left(\pa{V}{q^k} H^{i j k} - \frac{1}{m} g^{ij} - \frac{1}{m} g^{k l} p_l B^{i j}_k \right) dp_j \otimes \pa{}{q^i} + \left(\pa{V}{q^k} F^{j k}_i - \frac{1}{m} g^{k l} p_l D^j_{i k} \right) dp_j \otimes \pa{}{p_i} \end{multline*}
 
where the $A, B, C, D, E, F, G, H$ are the (for now arbitrary) connection potentials associated with the connection $\nabla$.
 
This means that

\begin{multline*} \Pi \circ \nabla(X_f) =\left(\pa{V}{q^k} G^{i k}_j - \frac{1}{m} \pa{g^{ik}}{q^j} p_k - \frac{1}{m} g^{k l} p_l A^i_{j k} \right) \pa{}{p_j} \otimes \pa{}{q^i} + \left(\pa{^2 V}{q^i \partial q^j} + \pa{V}{q^k} E^k_{i j} - \frac{1}{m} g^{kl} p_l G_{i j k} \right) \pa{}{p_j} \otimes \pa{}{p_i} \\
 + \left(\frac{1}{m} g^{ij} + \frac{1}{m} g^{k l} p_l B^{i j}_k - \pa{V}{q^k} H^{i j k} \right) \pa{}{q^j} \otimes \pa{}{q^i} + \left(\frac{1}{m} g^{k l} p_l D^j_{i k} - \pa{V}{q^k} F^{j k}_i \right) \pa{}{q^j} \otimes \pa{}{p_i} \end{multline*}

so that 

$$ d\pi \circ \Pi \circ \nabla(X_f) = \left(\frac{1}{m} g^{ij} + \frac{1}{m} g^{k l} p_l B^{i j}_k - \frac{1}{m} \pa{V}{q^k} H^{i j k} \right) \frac{\partial}{\partial q^j} \otimes \pa{}{q^i} $$

where $d \pi : TP \to TQ$ has been (uniquely) extended to the tensor product $TP \otimes TP$ bi-linearly.

Then 

$$\text{CSD} \circ  d\pi \circ \Pi \circ \nabla(X_f) = \left(\frac{1}{m} g^{ij} + \frac{1}{m} g^{k l} p_l B^{i j}_k - \frac{1}{m} \pa{V}{q^k} H^{i j k} \right) \left(\frac{\partial^2}{\partial q^j \partial q^i} - \Gamma^k_{i j} \frac{\partial}{\partial q^k} \right)$$

The first term here is exactly what we would like to see, while the extra two terms are extraneous from a physical perspective. So for this quantization procedure to be physically correct, we need $ B = H = 0$. In terms of the coordinate derivatives associated with a local fibered coordinate chart $\{ q^i, p_i \}$ on $P$, these connection potentials come from the terms

$$ \nabla_{\pa{}{p_j}} \pa{}{q^k} = B^{ij}_k \pa{}{q^i} + D^{j}_{ik} \pa{}{p_i} $$

and

$$ \nabla_{\pa{}{p_j}} \pa{}{p^k} = F^{j k}_i \pa{}{p_i} + H^{i j k} \pa{}{q^i} $$

The geometric meaning of both requirements is that covariant derivatives in the vertical directions should not induce changes in the horizontal directions.  

The coordinate invariant way to write the condition $B = H = 0$ while leaving all the other connection potentials arbitrary is to require that

\begin{equation} \label{connectioncondition} d \pi( \nabla_v u) = 0 \end{equation}

for all vertical vectors $v \in VP$ (that is, such that $v = v_i \pa{}{p_i}$ in local fibered coordinates) and all vector fields $u$ that satisfy $d \pi ([v, u]) = 0$ for all $v \in VP$. (Note that this second condition is in turn a coordinate invariant way of requiring that $\pa{u^i}{p_j} = 0$ for a vector field written in local fibered coordinates as $u^i \pa{}{q^i} + w_i \pa{}{p_i}$.)

Putting everything together -- including this new restriction on $\nabla$ -- we find that

\begin{equation} \label{ttqhamiltonian} Q_{TT} (H) = -  \frac{\hbar^2}{2m} \Delta + V(q^i) \end{equation}

where we have recognized $\Delta = g^{i j} \, \text{Hess}_{i j} = g^{i j} \frac{\partial^2}{\partial q^j \partial q^i} - g^{i j} \Gamma^k_{i j} \frac{\partial}{\partial q^k} $ as the Laplace-Beltrami operator associated with the metric $g$.
\end{proof}

Note that when we use global canonical coordinates paired with the canonical flat metric and connection, this gives us exactly the same result as in ordinary canonical quantization. It generalizes differently to non-flat metrics and their associated connections, but in a superior way: the general consensus is that the kinetic term in Hamiltonian operators should generalize to the Laplacian, not simple second derivative operators\cite{kleinert1997}, a result which here is naturally produced by the requirement of coordinate invariance.

\section{Conclusions} \label{conclusions}

In the previous section, I defined a quantization map that does precisely what I claimed in the introduction. By Theorem \ref{thm:ttq}, the map of \eqref{ttq}

\begin{itemize}
\item Mimics the major successes of canonical quantization 
\item Is mathematically well-defined and coordinate independent
\item Successfully quantizes spatial coordinate, momentum, angular momentum, and quadratic Hamiltonian functions
\end{itemize}

More specifically, \eqref{ttq} allows us to reproduce the results of canonical quantization in canonical coordinates on spaces with flat metrics. But it does this in a coordinate independent way, allowing us to carry out the quantization process in any coordinate system we like -- canonical or not -- and guarantee the same result. In other words, the following diagram commutes

$$
\begin{tikzcd}
\{ q^i, p_i, l_i, h \} \arrow[r,"Q_{TT}"] \arrow[d,"Y"] & \{Q(q^i), Q(p_i), Q(l_i), Q(H) \} \arrow[d,"Y"] \\
\{ Q^i, P_i, L_i, H \} \arrow[r,"Q_{TT}"] & \{ Q(Q^i), Q(P_i), Q(L_i), Q(H) \} \\
\end{tikzcd}
$$

Note that I have been careful here to specify precisely which phase space functions are successfully quantized in a coordinate independent way by \eqref{ttq}. I make no claim about the correct quantization of any other phase space functions. In particular, any phase space functions that do not neatly separate in terms of the sets $F_0(P)$, $F_1(P)$, and $F_2(P)$ are likely to be quantized in strange ways.

On the other hand, from a physical standpoint one might well argue that there are no other phase space functions beyond the ones listed above that we know how to quantize, as only phase space functions with known measurement properties -- that is, those whose quantum operators represent physical observables -- can be confirmed or rejected on physical grounds.

The first and most obvious objection to the tautologically-tuned quantization map of \eqref{ttq} is that it seems rather artificial. In particular, its only obvious virtue is to add quadratic Hamiltonians to the list of correctly quantized functions, and it does so at the cost of considerable additional complexity in comparison to the more elegant map of Kostant and Souriau. 

This objection is completely warranted as far as it goes, but it is unphysical. Physically, the phase space functions that need to be quantized are quite limited, and these physically important functions on phase space are now \emph{all} correctly quantized in a coordinate independent way. And this has been achieved without ever resorting to the esoteric techniques of geometric quantization: no complex polarizations, no half-form quantization, etc.

The only caveat here is that the construction of the tautologically-tuned quantization map of \eqref{ttq} requires an extra piece of mathematical structure beyond what is needed in ordinary geometric quantization, namely an affine connection on $TP$. Not only that, but it is necessary that this connection satisfy a particular condition, namely \eqref{connectioncondition}. So far, this extra structure -- and the associated condition -- seem to be necessary on physical grounds. This caveat would lose its teeth, however, if one could show that it is possible to construct an appropriate connection (that is, one that obeys the condition of \eqref{connectioncondition}) from some combination of the symplectic and Riemannian structures already given.

This is perhaps not so far fetched as it may sound. As one particular example of how this might work, consider that the Levi-Civita connection on $P$ induces an Ehresmann connection on $TP$, which in turn allows us to separate $TP$ into horizontal and vertical components. Since there are point-by-point isomorphisms between each of these horizontal spaces and the fibers of $TQ$ and between the vertical spaces and the fibers of $P$, we can therefore define a metric on $TP$ induced by the metrics on $TQ$ and $P$. Finally, there is a unique Levi-Civita connection compatible with this induced metric. So the structures already given can in fact be made to produce a connection on $TP$. However, the road is long and winding, and it is not yet clear if this process necessarily produces a metric that meets the requirements of Section \ref{ttquantization}. This type of construction is therefore a natural candidate for future work.

Even with the extra structure of an appropriate connection on $TP$, the tautological tuning process uses only symplectic and Riemannian structures that are present in both Hamiltonian particle systems and covariant Hamiltonian field systems. It therefore seems likely to be more readily extensible to the quantization of classical fields than traditional geometric quantization or deformation quantization have been. Indeed, some applications along these lines have already been tried, with partial success \cite{mcclain2025kostant}. This is therefore another natural candidate for future research in tautologically-tuned quantization.

\section*{Data availability statement}

Data sharing is not applicable to this article as no datasets were generated or analyzed for this analysis.

\section*{Funding} 

Partial funding for this work was received from the Washington and Lee Lenfest Grant program. There is no program ID associated with this program.

\section*{Competing interests}

The author has no competing interests to disclose relating to this analysis.

\bibliography{qss.bib}

\end{document}